\documentstyle[12pt]{article}
\input tcilatex
\QQQ{Language}{
American English
}

\begin{document} 

\title{General Birkhoff's Theorem }
\author{Amir H. Abbassi \\
 Department of Physics, School of Sciences,\\
Tarbiat Modarres University, P.O.Box 14155-4838,\\
Tehran, I.R.Iran\\
 E-mail: ahabbasi@net1cs.modares.ac.ir}
\date{}
\maketitle

\begin{abstract}
 Space-time is spherically symmetric if it admits the group of SO(3)
as a group of isometries, with the group orbits spacelike two-surfaces.
These orbits are necessarily two-surface of constant positive curvature.
One commonly chooses coordinate $\{ t, r, \theta,\phi\}$  so that the
group orbits become surfaces $\{ t, r= const\}$  and the radial coordinate 
$r$ is defined by the requirement that $ 4\pi r^2$ is the area of these 
spacelike two-surfaces with the range of zero to infinity. According to
the Birkhoff's theorem upon the above assumptions, Schwarzschild
metric is the only solution of the vacuum Einstein field equations. Our aim
is to reconsider the solution of the spherically symmetric vacuum Einstein 
field equations by regarding a weaker requirement. We admit the evident 
fact that in the completely empty space the radial coordinate $r$ may be defined so
that $4\pi r^2$ becomes the area of spacelike two-surfaces 
$\{ t, r=const\}$ with the range of zero to infinity. This is not necessarily
to be true in the presence of a material point mass M. It turns out that 
inspite of imposing asymptotically flatness and staticness as initial
conditions the equations have  general classes of solutions which Schwarzschild
metric is the only member of them which has an intrinsic singularity
at the location of the point mass M. The area of $\{t,r=const\}$
is $4\pi (r+\alpha M)^2$ in one class and $4\pi (r^2 +a_1 M r +
a_2 M^2)$ in the other class while the center of symmetry is at $r=0$.    

\end{abstract}

\newpage\
\begin{center}
\bf{Introduction}
\end{center}

Every spherically symmetric vacuum solution of Einstein's field equations
is part of the Schwarzschild solution according to the Birkhoff's theorem.
The solution possesses a four dimensional isometry group at least locally.
Historicaly Jebsen(1921) was the first to formulate it and Birkhoff(1923)
was the first to prove it\cite{1, 2, 3}. A popular proof has been presented
by Hawking and Ellis \cite{4}. Usually the method which applied to prove
this theorem is under a relatively stringent assumption. The radial coordinate
$r$ is defined by the requirement that $4\pi r^2$ is the area of the spacelike 
two-surfaces $\{ t,r=const\}$ which are invarient under the group of
isometries SO(3) operating on them. This assumption which is certainly
an evident fact in the completely empty space , physically is not necessarily always
to be true even in the presence of a material point mass particle M.
Our aim in this article is to reconsider the solution of the spherically 
symmetric vacuum Einstein's field equations i.e. Birkhoff's theorem
by regarding a weaker assumption. Let assume  only in completely empty space
radial coordinate $r$ may be defined so that $4\pi r^2$ to be the area
of spacelike two-surfaces with the range of zero to infinity. It results in
that Shwarzschild metric is not the unique solution but  general
classes of Schwarzschild form solutions exist which most of them are geodesically complete. The Schwarzschild metric
is the only member of them which possesses an intrinsic singularity at
the location of the point particle.\\
\begin{center}
\bf{Spherically Symmetric Space} 
 \end{center}

   Spherical symmetry requires the existence of a special coordinate system
$(t,r,\theta,\phi)$ so that in which the line element has the form
\cite{5, 6}

\begin{equation}  \label{1}
ds^2=B(r,t)dt^2-A(r,t)dr^2-2C(r,t)dtdr-D(r,t)(d\theta ^2+\sin {}^2\theta
\;d\varphi ^2)
\end{equation}

As mentioned above $r$ is defined so that the area of spacelike
two-surfaces $\{t,r=const\}$ becomes $4\pi r^2$.
There is a common belief that the line element can be transformed to the
following standard form by a suitable coordinate transformation. Some texts
start from here e.g.\cite{7}.

\begin{equation}     \label{2}
ds^2=B^{^{\prime }}(r^{^{\prime }},t^{^{\prime }})dt^{^{\prime
}2}-A^{^{\prime }}(r^{^{\prime }},t^{^{\prime }})dr^{^{\prime
}2}-r^{^{\prime }2}(d\theta ^{^{^{\prime }}2}+\sin {}^2\theta ^{^{\prime
}}\;d\varphi ^{^{\prime }2})
\end{equation}

Taking advantage of the mentioned form to compute the field equations it can
be shown that it is necessarily static and has a unique Schwarzschild
solution, as required by Birkhoff's theorem . This means that the
other solutions are just different forms of this metric which are related by
coordinate transformation. What indeed is flawing the reasoning, is that the
change of the coordinate $r\rightarrow r^\prime= \sqrt{D(r,t)}$ with new parameter
having the same range of $r$, while it is true in completely empty space
 is not necessarily to be generally true in the presence of material particle M.  Accordingly we believe that the steps
which follow to arrive at the standard form are not justified. Since it is a
hard task to solve the vacuum field equations with the general form of the
metric (1) we restrict our investigation to asymptotically flat and static
space-time by convention i.e.

\begin{equation}     \label{3}
ds^2=B(r)dt^2-A(r)dr^2-D(r)(d\theta ^2+\sin {}^2\theta \;d\varphi ^2)
\end{equation}
This metric tensor has the nonvanishing components

\begin{equation}     \label{4}
g_{tt}=-B(r)\;\;\;\;,\;\;\;\;g_{rr}=A(r)\;\;\;,\;\;\;g_{\theta \theta
}=D(r)\;\;\;,\;\;\;g_{\varphi \varphi }=D(r)\sin {}^2\theta
\end{equation}
with functions $A(r)\;,\;B(r)\;$and $D(r)$ that are to be determined by
solving the field equations. The nonvanishing contravariant components of
the metric are :

\begin{equation}      \label{5}
g^{tt}=-B^{-1}\;\;,\;\;\;g^{rr}=A^{-1}\;\;,\;\;g^{\theta \theta
}=D^{-1}\;\;,\;\;g^{\varphi \varphi }=D^{-1}\sin {}^{-2}\theta
\end{equation}

The metric connection can be computed by the use of (4) and (5) from the
usual definition. Its only nonvanishing components are:

\begin{equation} \label{6}
\begin{array}{llll}
\Gamma _{rr}^r=\frac{A^{^{\prime }}}{2A}\; & \Gamma _{\theta \theta }^r=- 
\frac{D^{^{\prime }}}{2A} & \Gamma _{\varphi \varphi }^r=-\sin {}^2\theta \; 
\frac{D^{^{\prime }}}{2A} & \Gamma _{tt}^r= \frac{B^{^{\prime }}}{2A} \\ 
\Gamma _{r\theta }^\theta =\frac{D^{^{\prime }} }{2D} & \Gamma _{\varphi
\varphi }^\theta =-\sin \theta \;\cos \theta &  &  \\ 
\Gamma _{r\varphi }^\varphi =\frac{D^{^{\prime }}}{2D} & \Gamma _{\varphi
\theta }^\varphi =\cot \theta &  &  \\ 
\Gamma _{tr}^t=\frac{B^{^{\prime }}}{2B} &  &  & 
\end{array}
\end{equation}
where primes stand for differentiation with respect to $r$. With these
connections the Ricci tensor can be obtained.

\begin{equation} \label{7}
R_{rr}=\frac{B^{^{\prime \prime }}}{2B}-\frac{B^{^{\prime }}}{4B}(\frac{%
A^{^{\prime }}}A+\frac{B^{^{\prime }}}B)-\frac{A^{^{\prime }}D^{^{\prime }}}{%
2AD}+\frac{D^{^{\prime \prime }}}D-\frac{D^{^{\prime }2}}{2D^2}
\end{equation}

\begin{equation}    \label{8}
R_{\theta \theta }=-1+\frac{D^{^{\prime }}}{4A}(-\frac{A^{^{\prime }}}A+ 
\frac{B^{^{\prime }}}B)+\frac{D^{^{\prime \prime }}}{2A}
\end{equation}

\[
R_{\varphi \varphi }=\sin {}^2\theta \;R_{\theta \theta } 
\]

\begin{equation}   \label{9}
R_{tt}=-\frac{B^{^{\prime \prime }}}{2A}+\frac{B^{^{\prime }}}{2A}(\frac{%
A^{^{\prime }}}A+\frac{B^{^{\prime }}}B)-\frac{B^{^{\prime }}D^{^{\prime }}}{%
2AD}
\end{equation}

\[
R_{\mu \nu }=0\;\;\;\;for\;\mu \neq \nu 
\]
The Einstein field equations for vacuum are $R_{\mu \nu }=0$. Dividing $%
R_{rr}$ by $A$ and $R_{tt}$ by $B$ and putting them together we get

\begin{equation}   \label{10}
-\frac{D^{^{\prime }}}{2AD}(\frac{A^{^{\prime }}}A+\frac{B^{^{\prime }}}%
B)+\frac 1A(\frac{D^{^{\prime \prime }}}D-\frac{D^{^{\prime }2}}{2D^2})=0
\end{equation}
Multiplying (10) by $\frac{2AD}{D^{^{\prime }}}$ gives

\begin{equation}     \label{11}
\frac{A^{^{\prime }}}A+\frac{B^{^{\prime }}}B=\frac{2D^{^{\prime \prime }}}{%
D^{^{\prime }}}-\frac{D^{^{\prime }}}D
\end{equation}
Now let integrate (11) with respect to $r$ and find

\begin{equation}  \label{12}
AB=C_1\frac{D^{^{\prime }2}}D
\end{equation}
where $C_1$is a constant of integration which can be fixed by requiring that 
$D$ asymptotically approaches to $r^2$ and $A$ and $B$ to one. This will fix 
$C_1$ to $\frac 14$ by (12), thus

\begin{equation} \label{13}
AB=\frac{D^{^{\prime }2}}{4D}
\end{equation}
Now using (9) and dividing the field equation $R_{tt}=0$ by $\frac{%
B^{^{\prime }}}{2A}$ we obtain

\begin{equation}    \label{14}
-\frac{B^{^{\prime \prime }}}{B^{^{\prime }}}+\frac 12(\frac{A^{^{\prime }}}%
A+\frac{B^{^{\prime }}}B)-\frac{D^{^{\prime }}}D=0
\end{equation}
Substituting (11) in (14) we get

\begin{equation}   \label{15}
\frac{B^{^{\prime \prime }}}{B^{^{\prime }}}+\frac{3D^{^{\prime }}}{2D}= 
\frac{D^{^{\prime \prime }}}{D^{^{\prime }}}
\end{equation}
The next step is to integrate (15) with respect to $r$ which gives

\begin{equation} \label{16}
B^{^{\prime }}D^{\frac 32}=C_2D^{^{\prime }}
\end{equation}
where $C_2\;$is a constant of integration. Dividing (16) by $D^{\frac 32}$
and taking another integration with respect to $r$ we get

\begin{equation}  \label{17}
B=-2C_2D^{-\frac 12}+C_3
\end{equation}
where C$_3$ is another constant of integration. $C_2$ and $C_3$ can be fixed
by considering the Newtonian limit of $B$ which gives $C_3=1$ and $C_2=M$ ($%
G=c=1$). Thus

\begin{equation}   \label{18}
B=1-2MD^{-\frac 12}
\end{equation}
$A$ may be computed by (12) and (18). It is

\begin{equation}    \label{19}
A=\frac{\frac{D^{^{\prime }2}}{4D}}{1-2MD^{-\frac 12}}
\end{equation}
It is a natural expectation that the functional form of $D$ to be fixed by
using the $\theta \theta $ component of the field equation, that is $%
R_{\theta \theta }=0.$ But using (18) , (19) and (11) it turns out that the
equation $R_{\theta \theta }=0$ will become an identical relation of zero
equal to zero for any analytic  
 function of $D$ which is only restricted to
the following constraints

\begin{equation}  \label{20}
D\rightarrow r^2\;\;\;\;\;\;if\;\;\;\;\;\;\;r\rightarrow \infty \;\cup
\;M\rightarrow 0
\end{equation}
This means that $D$ has the functional form

\begin{equation}  \label{21}
D(r,M)=r^2\;f(\frac Mr)
\end{equation}
where $f(0)=1$.  The appearance of square
root of $D$ in the final solution reveals that $D$ cannot adapt negative
values otherwise the metric becomes complex which physically is not
acceptable. Let's assume $D$ to be an analytic monotonic non-negative  increasing 
function of $r$ from zero to infinity and satisfies (21).There is no obligation that $D(0,M)$ becomes zero.
Since the functional form of  $f$  is not fixed then generally 
$D(0,M)$ may admit any positive value. Thus we may have the transformation
 $r\rightarrow r^{^{\prime }}=\sqrt{D(r,M)}%
=r\sqrt{f(\frac Mr)}$ but this does not specify the range of $r^{^{\prime }}$ 
at all. It is clear that the Schwarzschild solution is a special case of the general
solution by choosing $\;f=1,$ indeed the simplest case. This does not based
on any physical fact. This unjustified choice which is not even stand on any
fundamental reasoning, is the source of the trouble of singularity in some
parts of space-time and of course can be avoided. Then we may perform the following coordinate transformation

\begin{equation}    \label{22}
r\;\text{with the range }(0,+\infty )\;\rightarrow \;r^{^{\prime
}}=D(r,M)^{\frac 12}\;\text{with the range }(D(0,M)^{\frac 12},\infty)
\end{equation}
where $D(0,M)$ may be any arbitrarily real non-negative number. Since $D$
has $\left[ L^2\right] $ dimension and $D(0,0)$ is equal to zero and also $M$
is the only natural parameter of the system which has length dimension we
may conclude that the general form of $D(0,M)$ is

\begin{equation}   \label{23}
D(0,M)^{\frac 12}\;=\;\alpha (M)M
\end{equation}
where $\alpha $ is a dimensionless parameter which for simplicity may take
it as a constant independent of $M$. Now we make another coordinate
transformation

\begin{equation}    \label{24}
r^{^{\prime }}\;\rightarrow \;r^{^{\prime \prime }}=r^{^{\prime }}-\alpha M
\end{equation}
The range of new radial coordinate is from zero to infinity. Dropping primes
the final form of the metric becomes

\begin{equation}   \label{25}
ds^2=(1-\frac{2M}{r+\alpha M})dt^2-\frac{dr^2}{1-\frac{2M}{r+\alpha M}}%
-(r+\alpha M)^2(d\theta ^2+\sin ^2 \theta d\varphi ^2)
\end{equation}
Thus $\alpha $ is an arbitrary constant which its different values define
the members of our general solutions family \cite{8,9}. The only condition on $\alpha $
is that it should not be much much bigger than one, otherwise it would
contradict with Newtonian mechanics predictions. The corresponding metrics
related to the values of $\alpha $ bigger than 2 are all regular in the
whole space. The actual metric of the spacetime of course is merely a
special member of this class but not necessarily the Schwarzschild metric($%
\alpha =0$). Specifying this requires further information about the actual
properties of spacetime at Schwarzschild scales which at this time there is
no access to such data.

\begin{center}
\textbf{Test for Equivalence}
\end{center}

Someone may be taken in by the apparent form of the general solution that
these are just the usual Schwarzschild solution with $\;r$ \ replaced by $%
\;r+\alpha M$ and conclude that they are not new. In the following we
clarify this point and show that it is not so.

\noindent 1- In Schwarzschild solution the range of radial coordinate $r$ is
between $0\;$to$\;\infty $ and the center of symmetry is at $r=0$. Replacing 
$r$ by $\acute{r}+\alpha M$ does not lead to the new solution because the
range of the transformed radial coordinate is between $-\alpha M$ \ to $%
\infty .$ Of course negative values for conventional radial coordinate is
meaningless and the center of symmetry is located at $\acute{r}=-\alpha M$.
This is not identical to the new solution because the range of radial
coordinate is between $0$ to $\infty $ and the center of symmetry is at $%
\acute{r}=0$. On the other hand in the new metric replacing $r+\alpha M= 
\acute{r}$ does not lead to Schwarzschild solution because the range of $%
\;r\;$here is between $0$ to $\infty $ and the center of symmetry is at $r=0$%
. While the range of the transformed radial coordinate is between $\alpha M$
to $\infty $ and the center of symmetry is at $\acute{r}=\alpha M$. This
evidently is different from Schwarzschild solution. Though the extension of $%
\acute{r}$ to values smaller than $\alpha M$ is physically meaningless
because a point by definition has no internal structure to be extended
inside of it, let us consider such a mathematical hypothetical spacetimes.
Even these are essentially different from Schwarzschild spacetime because in
Schwarzschild the point mass $M$ is at $r=0$ while in these spacetimes it is
located at $r=\alpha M$.

\noindent 2- The center of spherical symmetry that is the position of point
mass $M$ is a common point between the Schwarzschild spacetime and the
presented spherically symmetric vacuum spacetimes in this manuscript. As it
has been shown the field equations and the given boundary conditions are not
sufficient to fix $\alpha $. Thus if we take $\alpha \neq 0$ , then these
solutions will not be singular at the center of symmetry while the
Schwarzschild spacetime possesses an intrinsic singularity at the center of
symmetry. If these new metrics were isometric to Schwarzschild metric they
should be singular too, because coordinate transformation cannot change the
intrinsic properties of spacetime. This clearly shows that the presented
metrics are not Kottler solutions of Schwarzschild spacetime.

\noindent 3- The presented general solution and the Schwarzschild solution
have exactly the same space extension. Making use of Cartesian coordinate
system as frame of reference will elucidate this fact. It turns out that all
components have the same range $(-\infty ,+\infty ).$

\noindent 4- Let us consider hypothetically spacetimes which possesses
different lower bound for the surface area of a sphere. Obviously they have
different geometrical structures and present different physics.

\noindent 5- The zone of $r$ of the order of Schwarzschild radius is the
domain in which gravitational field is tremendously strong and
conventionally we have to give up our common sense and replace the character
of $r$ and $t$. So it is not surprising to have a geometry completely
different.

\noindent 6- Schwarzschild solution is a special case of our general solutions 
Eq.(25) for the case of $\alpha=0$. Therefore both Schwarzschild and 
general solutions are in the same coordinate system which manifestly
have different form. Any transformation to new radial coordinate
$r^{^\prime}=r+\alpha m$ requires the Schwarzschild metric be 
written in this new coordinate too which means we should replace 
$r$ by $r^{^\prime}-\alpha m$. Thus in the new coordinate they
will have different form too. This means general solutions are not
isometric to any piece of the Schwarzschild metric.\\
\begin{center}
\bf{Completeness}
\end{center}
For $\alpha>2$ in (25) $t$ remains time coordinate everywhere.
This means the hypersurfaces $t=const.$ become spacelike positive
definite Riemanian 3-dimensional manifolds. These are metrically
complete because every Cauchy sequence with respect to the distance
function converges to a point in the manifold. It is well known that
metric completeness and geodesically completeness are equivalant for a positive definite metric \cite{10}.\\
\begin{center}
\bf{The Most General Case}
\end{center}

What we have assumed in the previous sections for $D(r,M)$ i.e. to be 
an analytic monotonic non-negative increasing function is not generally a
necessary requirement. It is merely necessary to be analytic and non-negative.
To see how this may be possible let us consider a series expansion of
$f(\frac Mr)$ as follows;
\begin{equation}
f(\frac Mr)=\sum\limits_{n=-\infty}^{+\infty} a_n (\frac Mr)^n
\end{equation} \label{26}
The condition, $f(0)=1$ implies that 
\begin{eqnarray}
a_n &=&0 , \;\;\;\;\; n=-1, -2, ...\; , -\infty \nonumber \\
a_0 &=&1
\end{eqnarray}
Since $D(r,M)$ is supposed to be analytic everywhere including $r=0$,
this leads to 
\begin{equation}
a_n =0 \;\;\;, \;\;\; n=3, 4,\;...\; , +\infty
\end{equation} \label{28}
Thus the most general form of $D(r,M)$ is 
\begin{equation}
D(r,M)=r^2 f(\frac Mr) = r^2 +a_1 Mr +a_2 M^2
\end{equation} \label{29}
where $a_1$ and $a_2$ are dimensionless numbers which should be
fixed by comparision with observation. 
The non-negativeness of $D(r,M)$ would be guaranteed if 
\begin{eqnarray}
a_2 \geq 0 \;\;\; &and&\;\;\; {a_1}^2 -4a_2 \leq 0 \nonumber\\
&or&  \;\;\; -2\sqrt{a_2} \leq  a_1 \leq  2\sqrt{a_2}
\end{eqnarray}  \label{30}
The minimum value of $D(r,M)$ is at 
\begin{equation}
r=-\frac{a_1 M}2
\end{equation}  \label{31}
For $a_1 \geq 0$  (29) is a non-negative monotonic and increasing function of $r$ 
from zero to infinity as the same case which was discussed previously.
But for $ -2\sqrt{a_2} \leq a_1 \leq 0$ ,$ \;$ $D(r,M)$ decreases as $r$ goes 
from $+\infty$ to $-\frac{a_1 M}2$ and then increases as $r$ goes 
from $-\frac{a_1 M}2$ to zero. Thus for the case $a_2 \geq 0$ and 
$-2\sqrt{a_2}\leq a_1\leq 0$ it is not possible to have a transformation like
$r \rightarrow r^{\prime}=\sqrt{D}$. Becuase there is not  one to one
correspondence between $r$ and $r^{\prime}$ in this case.\\
We may write the most general form of the solution as 
\begin{eqnarray}
ds^2&=& (1-2MD^{-\frac12})dt^2 - \frac{\frac{D^{^{\prime }2}}{4D}}{1-2MD^{-\frac 12}}dr^2 -D(d\theta^2 + sin^2\theta
 d\phi^2) \\
D&=&r^2 +a_1 Mr +a_2 M^2 \;\;\;and\;\;\;
a_2\geq 0 \;\; , \;\;\; -2\sqrt{a_2}\leq a_1\leq 0 \nonumber
\end{eqnarray} \label{32}
This metric would be free of any coordinate singularity if we have 
\begin{equation}
a_2 -\frac{{a_1}^2}4 >4
\end{equation}  \label{33}
(33) will restrict the allowed values of $a_1$ and $a_2$ to the range 
\begin{equation}
a_2 >4 \;\;\; and \;\;\; -2\sqrt{a_2 -4}<a_1<0
\end{equation} \label{34}
The proof of completeness which was presented for (25) does hold 
for (32).

\begin{center}
\bf{Conclusion}
\end{center} 

We may conclude the discussion  that 
Birkhoff's theorem to be modified with this statement that
 the vacuum Einstein field equation spherical solutions  are uniqely
of general Schwarzschild-like form (25) or (32) which may be called
general Birkhoff's theorem. Spacetimes (25) for $\alpha>2$  and (32)
for (34) are nonsigular and maximally extended.


\begin{thebibliography}{1}
\bibitem{1} Schmidt,J.H , gr-qc/97071.

\bibitem{2} Jebsen, Ark.Mat.Astron.Fys. 15, 18 (1921).

\bibitem{3}  Birkhoff,G.D. \emph{Relativity and Modern Physics}, Harvard
Univ.Press(1923).

\bibitem{4} Hawking, S.W. and  Ellis, G.F.R. \emph{ The Large Scale
 Structure of Space-time}, Cambridge University press,(1973).

\bibitem{5}  Weinberg,S. \emph{Gravitation and Cosmology }, John
Wiley,p335(1972).

\bibitem{6}  D'Inverno,R. \emph{Introducing Einstein's Relativity},
Clarendon Press,p185 (1992).

\bibitem{7}  Rindler,W.\emph{Essential Relativity },Springer-Verlag,
p136(1977).

\bibitem{8}  Abbassi, Amir H. ,  gr-qc/9812081.

\bibitem{9}  Gharanfoli, S. and  Abbassi, Amir H. ,  gr-qc/9906049.

\bibitem{10}  Kobayashi, S. , Nomizu, K. \emph{Foundations of Differential
Geometry: Volume I },Interscience,New York, (1963).

\end{thebibliography}
\end{document}